\documentclass[11pt]{article}
\usepackage[tbtags]{amsmath}
\usepackage{graphicx}
\usepackage{amssymb}
\usepackage{epstopdf}
\usepackage{natbib}
\usepackage{citecollapse}
\title{Robust Monomer-Distribution Biosignatures in Evolving Digital Biota}
\author{Evan D. Dorn$^{1}$, Christoph Adami$^{1,2,3,4}$}
\begin{document}

\maketitle
{\it$^{1}$Computation and Neural Systems 139-74, California Institute of Technology \\*
$^{2}$Keck Graduate Institute of Applied Life Sciences, Claremont, CA 91711\\
$^{3}$Department of Microbiology and Molecular Genetics, Michigan State University\\
$^{4}$The BEACON Center for the Study of Evolution in Action, Michigan State University }

\begin{abstract}
Because organisms synthesize component molecules at rates that reflect
those molecules' adaptive utility, we expect a population of biota to
leave a distinctive chemical signature on their environment that is
anomalous given the local (abiotic) chemistry.  We observe the same effect in the
distribution of computer instructions used by an evolving population of
digital organisms, and characterize the robustness of the evolved
signature with respect to a number of different changes in the system's
physics.  The observed instruction abundance anomaly has features that
are consistent over a large number of evolutionary trials and
alterations in system parameters, which makes it a candidate for a
non-Earth-centric life-diagnostic. \end{abstract}

\section*{Introduction}
When searching for signatures of extraterrestrial life, one is inevitably drawn into the quandary of extrapolating from terran biochemistry to 
universal principles~\citep{WardBenner2007}. If we stipulate that all forms of life must be chemistry-based, and encode information about their world within molecules that evolve according to Darwinian rules, then the set of possible chemistries, while still unlimited, is constrained by universal features~\citep{BennerSwitzer1999,Bains2004,Benner2010}. In particular, 
evolved bio-organisms impress a distinctive chemical signature on their
environment, because biota synthesize those compounds that are necessary for
competition and replication.  More generally, biochemical synthesis may
be seen as the product of natural selection, as evolution shapes genomes so as
to maximize their fitness.   As a result, the chemical species that
persist in a biotic environment generally deviate from the chemical species
we would expect in the absence of life.  

Previously, we demonstrated~\citep{DornED_etal:2011} that a quantifiable phenomenon
(which we call the Monomer Abundance Distribution Biosignature or MADB) is measurable in both 
terrestrial biochemicals and in the artificial life system Avida.   Patterns of monomer 
concentration (amino acids or carboxylic acids in the biosphere, or computer 
instructions in the digital life environment) reliably distinguish between life-bearing
and sterile environments.  The biotic patterns appear to reflect evolutionary constraints
on the organisms' composition and function, while abiotic patterns reflect thermodynamic and
formation-kinetic constraints. The evolutionary constraints are many: in terrestrian biology for example, proteins must be specific to achieve particular functions given other proteins and small molecules they interact with. But they must also fold reliably and if they are soluble in water they must have a hydrophobic core. Both functional and structural constraints affect which amino acids appear in a sequence~\citep{Forsdyke2005}, and the same is expected for the composition of programs in artificial life. 

This study examines the robustness of the MADB, and attempts to demonstrate with a higher
degree of confidence that it results from selection pressures acting on evolving biota.  A robust
biosignature is one that manifests reliably in the presence of life.  We
examine this hypothesis by altering the underlying ``thermodynamics"
of the artificial life system and measuring the robustness of this biosignature
as the digital biota evolve under varying conditions.   Our goal is not just to demonstrate 
that the biosignature forms but to measure the extent to which its features are the result of 
selection by quantifying their independence from the underlying physics.

As we will see, the MADB is largely, but not entirely, conserved
even as the physics are changed.  Examination of which aspects are conserved can 
be linked to understanding the function, behavior, and adaptive needs of the digital
biota.

\section*{The Monomer Abundance Distribution Biosignature (MADB)}

The rates of formation and diagenesis of individual chemical monomers
(e.g. amino acids, carboxylic acids, or other ligands) in the absence of
life are dictated by the laws of formation kinetics and thermodynamics.
Consequently, the observed relative abundances of various monomers in an
abiotic environment reflect these constraints.  For example,
when amino acids are formed without life, large and thermodynamically
expensive molecules such as valine are always seen at drastically lower
concentrations than simpler compounds like glycine and alanine~\citep{DornED_etal:2003,KvenvoldenKA_etal:1971,McDonaldGD_etal:1994,MillerSL:1953,MillerSL:1957,Munoz-CaroGM_etal:2002,WolmanY_etal:1972}. 

Organisms, on the other hand, are constrained by their need to reproduce
and compete.  Biota expend energy to manufacture whatever monomers are
necessary to meet a fitness criterion: while synthesizing a particular
molecule may be relatively expensive, if it is essential to competition
the alternative may be extinction.  Therefore, we expect evolved
genotypes to synthesize molecules at rates that reflect those molecules'
utility in fitness rather than, or in addition to, their
thermodynamic cost.     This effect has been
previously discussed by \cite{Lovelock:1965,McKay:2002,McKay:2004,SummonsRE_etal:2008,ShapiroR_Schulze-MakuchD:2009,DaviesPCW_etal:2009}.  McKay in \citep{McKay:2004} and \citep{DaviesPCW_etal:2009} coined the term ``Lego Principle" to
describe the specific case that biological systems employ a discontinuous subset of the possible molecules
in a family of biochemicals, for example that terrestrial biota use only two dozen or so 
out of a much larger number of possible amino acids. Such a signature has also been proposed as a means to discover a ``shadow biosphere" on Earth: a form of biochemical life with an independent origin from the life we know today~\citep{DaviesPCW_etal:2009}. 

We call any measurable variation between biotic and abiotic monomer concentrations
within a chemical family the ``monomer abundance distribution biosignature" (MADB).
The MADB is very pronounced in the terrestrial biosphere, and is easily
detectable by a number of mathematical techniques~\citep{DornED_etal:2003,DornED_etal:2011}.

\section*{Artificial Life as a Testbed for Biosignatures}

One of the fundamental concerns for any putative biosignature is that we have only
one biosphere to test it against.    Yet an ideal biosignature
should be capable of detecting life regardless of its particular biochemistry.  Moreover,
we cannot conduct experiments to ``start evolution over" in the terrestrial biosphere 
and examine the resulting evolved biochemistry. To overcome these hurdles, we turn to artificial life (in particular {\em digital life}), for an additional example of life and of
evolutionary processes that we can use as a testbed.   Because life {\em in silico} is
unrelated to terrestrial biochemistry, this also serves to help abstract our thinking 
and avoid assumptions resulting from a terrestrial bias.

Here, we use the artificial life platform ``Avida"  (an introduction to Avida may 
be found in \cite{OfriaC_WilkeCO:2004}, see~\cite{Adami2006} for a review of 
research performed with artificial life techniques).   Avida organisms
(``avidians") are small, self-replicating programs written in a simple
programming language; 29 instructions are available in the
variant used for this study.  The instructions may be seen as
analogous to the monomers (such as amino acids) that compose familiar
biota because the frequency with which instructions appear within the avidian genomes is a very good proxy for the frequency with which they are executed, that is, the phenotype of the organism. This is due to the linearity of program execution in Avida, where loops are uncommon (except for a single, usually short, loop used for replication).  While the instructions/amino acids analogy is inexact, Avida instructions share three key properties with chemical monomers:  the ``biomass'' of avidians is composed of those
instructions, and that composition is both inheritable and subject to 
selection.  If the constituents' (whether biochemical or digital) relative concentrations
are measurable, these properties are all that is necessary for an MADB to form. ,

In Avida, instructions are substituted and inserted into genomes via 
externally-imposed mutations including copy errors, point mutations, and insert
and delete mutations.  Genomes have multiple options of
monomers from which to construct genes, and by default, all instructions
appear with equal probability when a mutation is imposed.  In this world,  this represents a
the fundamental abiotic process, since avidians cannot affect the 
mutation rate or the frequency of appearance of any particular
instruction.   If adaptation did not constrain their abundance, 
we would expect all 29 instructions to appear in equal proportion 
in the population.

\begin{figure}[htbp] %  figure placement: here, top, bottom, or page
   \centering
   \includegraphics[width=5in]{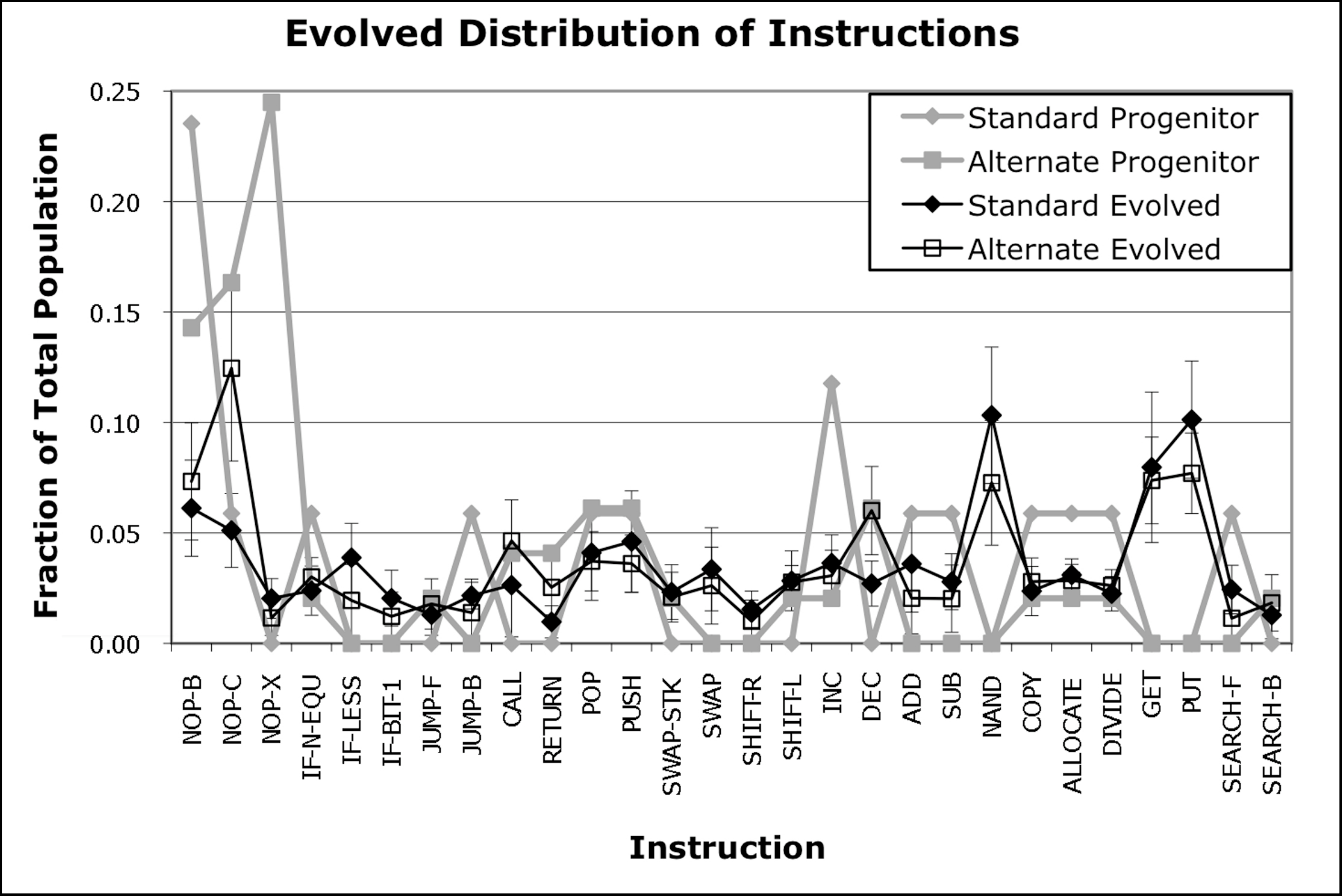} 
   \caption{The distribution of instructions in two different ancestor
organisms and in populations descended from those ancestors.  While the
ancestors have very different composition, the composition of the descendants is similar
as the terminal populations have adapted to the same environment.  ``Evolved" lines
represent the average of 25 different evolutionary trials, each sampled after 1500 
generations.  Error bars are one standard deviation.  (Data from the experiment 
described in~\protect\citealt{DornED_etal:2011}, in which all instructions were equally
available in mutation). There are no errors for the progenitor distributions as they are exact.} 
   \label{fig:fig1}
\end{figure}

When the bulk frequency of programming instructions is counted
in evolved populations in Avida, we observe a distinct profile that
does not reflect either the abiotic parameters of the system or the
instruction frequency of the ancestor, indicating that selection has 
dictated the monomer abundance pattern of the population.   This pattern is largely
consistent over many trials, even though the actual genomes evolved may
not resemble each other at all.  Moreover, the pattern is
consistent over a wide array of parameters such as mutation rate
and different ancestors.  Figure~\ref{fig:fig1} shows the relative distribution
of 28 computer instructions in evolved Avida populations that are descendant from
two distinct ancestor genotypes (these results are more fully presented
in \citealt{DornED_etal:2011}).  One instruction (NOP-A) is excluded from
our analysis; see Methods for a discussion.  Note that while the ancestors have very
different composition, their descendants have converged to some extent to a common
profile, demonstrating the dominant effect of selection on monomer abundances.

\begin{figure}[htbp] %  figure placement: here, top, bottom, or page
   \centering
   \includegraphics[width=5in]{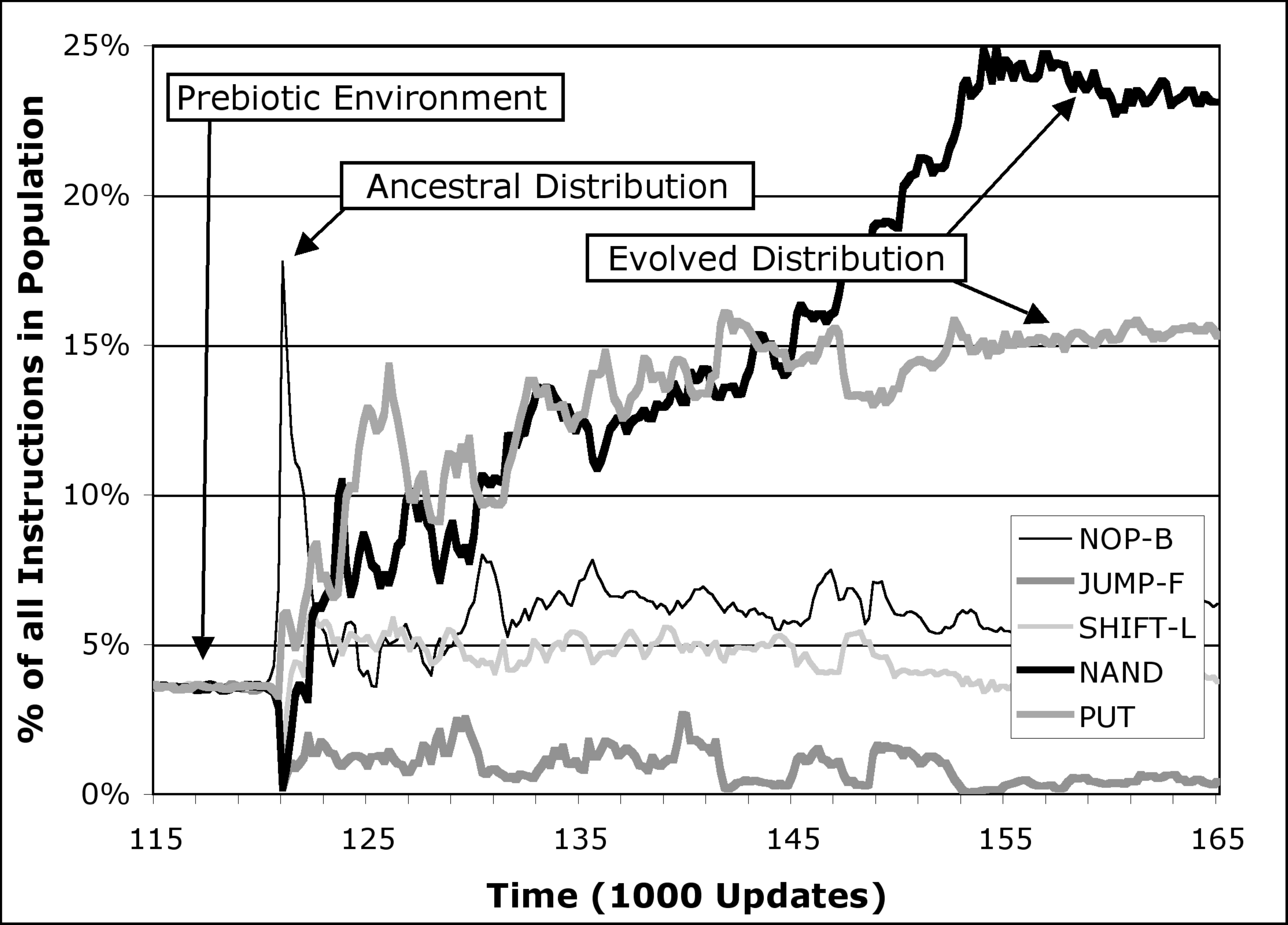} 
   \caption{Evolution of the distribution of six computer instructions as incident
 self-replicators colonize and adapt to a a formerly lifeless environment.  At the
 outset, mutations cause all instructions to be present in roughly equal proportion.
 The ancestor organism's genome, nearly 20\% NOP-B, dominates the early biotic
 distribution.  As the organisms adapt to the environment, a NAND-heavy distribution
 develops and stabilizes.  Often-lethal instructions such as JUMP-F are strongly suppressed
 by selection pressure. From Dorn et al.\ (2011).}
   \label{fig:fig2}
\end{figure}

When life is introduced to a formerly abiotic environment,  the MADB rapidly 
overwhelms the preexisting abiotic signature, as seen in Fig.~\ref{fig:fig2}. In the experiment that produced Fig.~\ref{fig:fig2}, an Avida population was seeded with randomly generated, nonviable genomes and bombarded with a high (lethal) level of point
mutations.  Single, viable intact organisms were periodically introduced
into the environment while the rate of point mutations was stepped down.
 When the mutation rate became low enough for organisms to survive,
avidians quickly populated the entire landscape, impressing their
signature distribution of instructions onto the environment.  An initial
spike reflects the ratios of instructions present in the ancestor
genotype, but this was quickly replaced by an evolved MADB as the
organisms adapted.

In this study, we further explore the robustness of the MADB as the fundamental abiotic
parameters of the Avida environment are changed.  This is important because it could be
argued that the MADB observed in terrestrial biochemicals (e.g., amino
acids) is highly dependent on the formation thermodynamics of the
individual monomers, and that the pattern would be drastically altered
if the costs of synthesis were changed.  Alternatively, it is
conceivable that the distinction seen between biotic and abiotic patterns
are not the product of selection, but of some other, unknown function. 
In artificial life, we can test these conjectures.  

Specifically, to study the robustness of the MADB in digital organisms, we alter
the availability of each instruction by changing the frequency with
which it appears in mutation:  this is loosely analogous to altering the
formation thermodynamics of amino acids, thus changing their
availability to early life forms, or else to alter the mutational bias on individual nucleotides.  If elements of the MADB pattern are
retained despite these alterations, it demonstrates that selection is
capable of overwhelming the constraints of physics with respect to the
composition of organisms in early evolution.

We hypothesize that some instructions' presence (or absence) will be
more or less independent of the frequency with which they appear in
mutation, indicating that their appearance frequency in the genome is
strongly constrained by a fitness criterion, while other instructions
are less strongly constrained.  Instructions that convey a strong
fitness benefit should be incorporated into genomes rapidly, thus
ensuring that they account for a large proportion of the final
population. Anti-adaptive instructions (i.e., ones that are more often
deleterious when appearing as mutations) should be suppressed in the
population. We should emphasize that not all deviations from the frequency with which an instruction is created by mutations is adaptive. In any evolutionary unfolding, changes can be due to chance, due to adaptation, or due to historical contingency~\citep{Travisanoetal1995,Joshietal2003,WagenaarAdami2004}. For example, one instruction that could be used as an alternative to another might be ``locked in" early during evolution and appear at an increased frequency throughout history when the alternative could just as well have been used. At the same time, neutral drift and chance events could affect instruction frequency, even though such a frequency would not be stable. Because all these effects are expected to shape the frequency distribution of any system undergoing Darwinian evolution, they do not distract from the universality of the MADB.  

\section*{Methods}

By default, Avida substitutes new instructions (monomer synthesis) during mutation events
with an equal probability for each instruction.  However, this is unlikely to be realistic in monomer chemistry as each monomer can be expected to have its own formation probability. Thus, this ``probability of synthesis" should have a bias
that reflects the environment's physics or chemistry. We model different such biases 
by constructing systematic biases by hand, or else by creating random biases. For this experiment we
created a modified version of Avida version 1.6 that allows the experimenter to
specify a probability of substitution for each instruction (in the standard version, each instruction is substituted with equal probability), and which includes a nonstandard output to report the population frequency of each instruction.
The code for this version of Avida, along with the configuration
files used, are available in the online supplemental information for this article
or from the authors.

\subsubsection*{Experiments A-D (Figure 3)} In each experiment, a grid of 3600 cells was populated with a
13-instruction simple self-replicating ancestor, whose length can change due to insertion and deletion mutations.  This initial
population was evolved for 1500 generations and the bulk frequency of
each instruction in the population was quantified every 100 generations. The bulk frequency of an instruction is given by the total number of
that instruction in the population, divided by the total number of instructions in the population (3600 times average sequence length). The run time of 1500 generations was chosen based on preliminary experiments (not shown) that demonstrated that even with evolution 
still ongoing, the MADB  is reliably established in the first few hundred generations and tends not to change extensively after that time.  
To provide smoothing of momentary fluctuations in instruction concentration, the frequency of each of the 29 instructions was measured during each of the last ten generations, and averaged over these.  We tested four different distributions of substitution probabilities: an increasing distribution (Experiment A), a decreasing distribution (Experiment B), one that increases and then decreases (Experiment C), as well as a randomly generated probability distribution (Experiment D). Note that as the order of instructions is arbitrary, these four different substitution patterns have no inherent meaning. We performed 25 replicates of each experiment. All other parameters used in these experiments are the standard defaults described in~\citep{OfriaC_WilkeCO:2004}, except that the standard fitness landscape of nine logic tasks (all distinct one- and two-input tasks) was replaced with the extended 73 logic tasks landscape, where all distinct logic tasks with up to three inputs are rewarded.

\subsubsection*{Mutation ``spectra" experiments (Figure 4)} We created eighteen different ``spectra" that represent the relative
frequency with which each instruction appears through mutation. For this experiment, we used the three  manually constructed systematic variations of experiments A-C, while another fifteen experiments used randomly-generated mutation spectra, where each instruction was assigned a substitution probability bound between 0 and 0.08. This implies that any instruction that is assigned a vanishing substitution probability can never appear in the population. Each experiment was performed in 25 replicates. 

\subsubsection*{Analysis} In our data analysis, we consider only 28 of the 29 instructions used in
these Avida populations.  One instruction, NOP-A, is used by the system
as a temporary placeholder to initialize empty memory in dividing organisms,
but is replaced as the organism copies its actual genotype into the empty array.  
Therefore, the number of NOP-A instructions appearing in the population 
fluctuates rapidly in a way that has no biological analog and is highly dependent
on the precise timing of the sample measurement.  Also, this aspect of NOP-A
cannot be selected for or against since the organisms do not have a choice 
about how empty memory is initialized.  We therefore choose to exclude 
it from the analysis.

\section*{Results}

\begin{figure}[htbp] %  figure placement: here, top, bottom, or page
   \centering
   \includegraphics[width=3.25in]{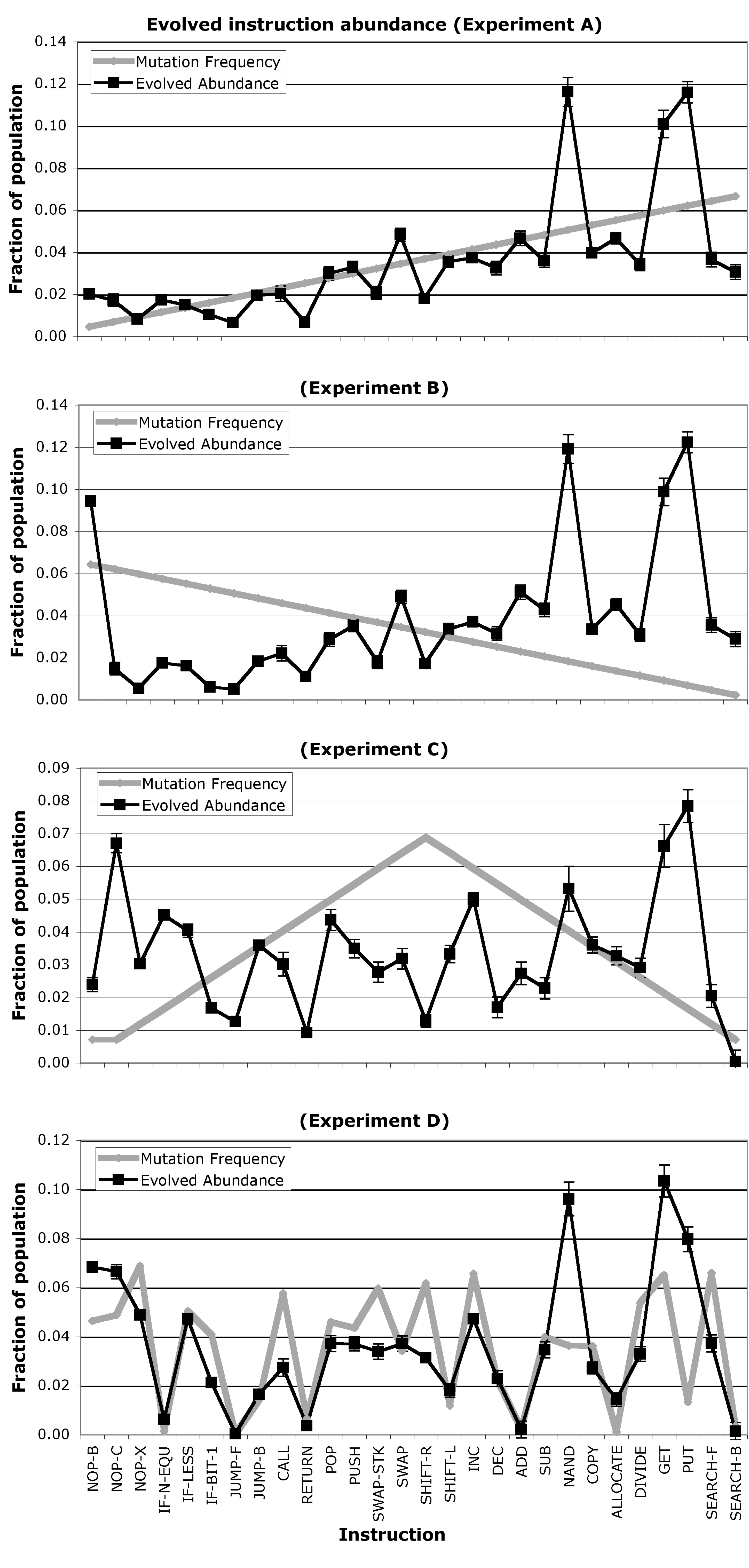} 
   \caption{Four experiments showing the evolved distribution of instructions 
as the underlying physics are changed.  The gray curves represent the frequencies
with which each instruction was presented to organisms through mutation, black curves
represent the relative abundances of the instructions in evolved populations.
Each black curve represents the average of 25 populations, error bars are standard error.
The general features of the selection-driven distribution are conserved even though the 
mutation frequencies are varied over large ranges.  In experiments A, B, and C, the most
frequent instructions appear 30 times more often than the least-frequent.  Experiment D
shows one of 15 randomly-generated distributions tested.}
   \label{fig:fig3}
\end{figure}

Evolved Avida populations impress a distinctive pattern of instruction
abundances onto their environment, and this pattern is largely conserved
even when the availability of particular instructions is altered
significantly.  Fig.~\ref{fig:fig3} shows the average evolved abundances of the 28
instructions for the four different mutational profiles described in Methods.  In each, the gray
``mutational frequency" line represents the relative rates at which each
instruction appears in mutation, and the black ``evolved frequency" line
represents the relative abundance of each instruction in the terminal
population.

Certain features, such as the prominence of GET, PUT, and NAND are
conserved across the runs regardless of how their input (mutational)
frequency is altered.  This reflects the fitness benefit conveyed by
these instructions, which are necessary for completing logic computations. In
Avida, organisms are rewarded with increased processor time for
successfully completing a variety of computational tasks, which play the
role of exothermic catalytic reactions in the metabolism of digital
organisms~\citep{Adami2006}.  In the landscape we used, 73 tasks are rewarded, each of which requires one to five NAND operations.  
Some other instructions, such as JUMP-F and RETURN, are frequently lethal when a mutation causes
them to appear in the genome.   As a result, they are rarely incorporated 
into genomes and appear underrepresented in the final population regardless of their mutational
frequency. In their effect, such instructions are not unlike DNA codons that cause early termination of transcription in terrestrial biochemistry. 
A comparison between the four experimental treatments shows that the adaptive component dominates the mutational bias. Yet, the influences of chance and history~\citep{Travisanoetal1995,WagenaarAdami2004} are also present. For example, the relative abundance of the modifier instructions NOP-B and NOP-C differ in the different treatments, even though they could in principle substitute for each other. Even though the roles of the two instructions are similar, each instruction is assigned a particular functional role early on, and can only be released from this role in very unlikely mutational events. This historical contingency is at the origin of many run-to-run differences between instruction abundances. But as long as the abundances differ significantly from the abiotic baseline, they still contribute to the MADB.

\begin{figure}[htbp] %  figure placement: here, top, bottom, or page
   \centering
   \includegraphics[width=6in]{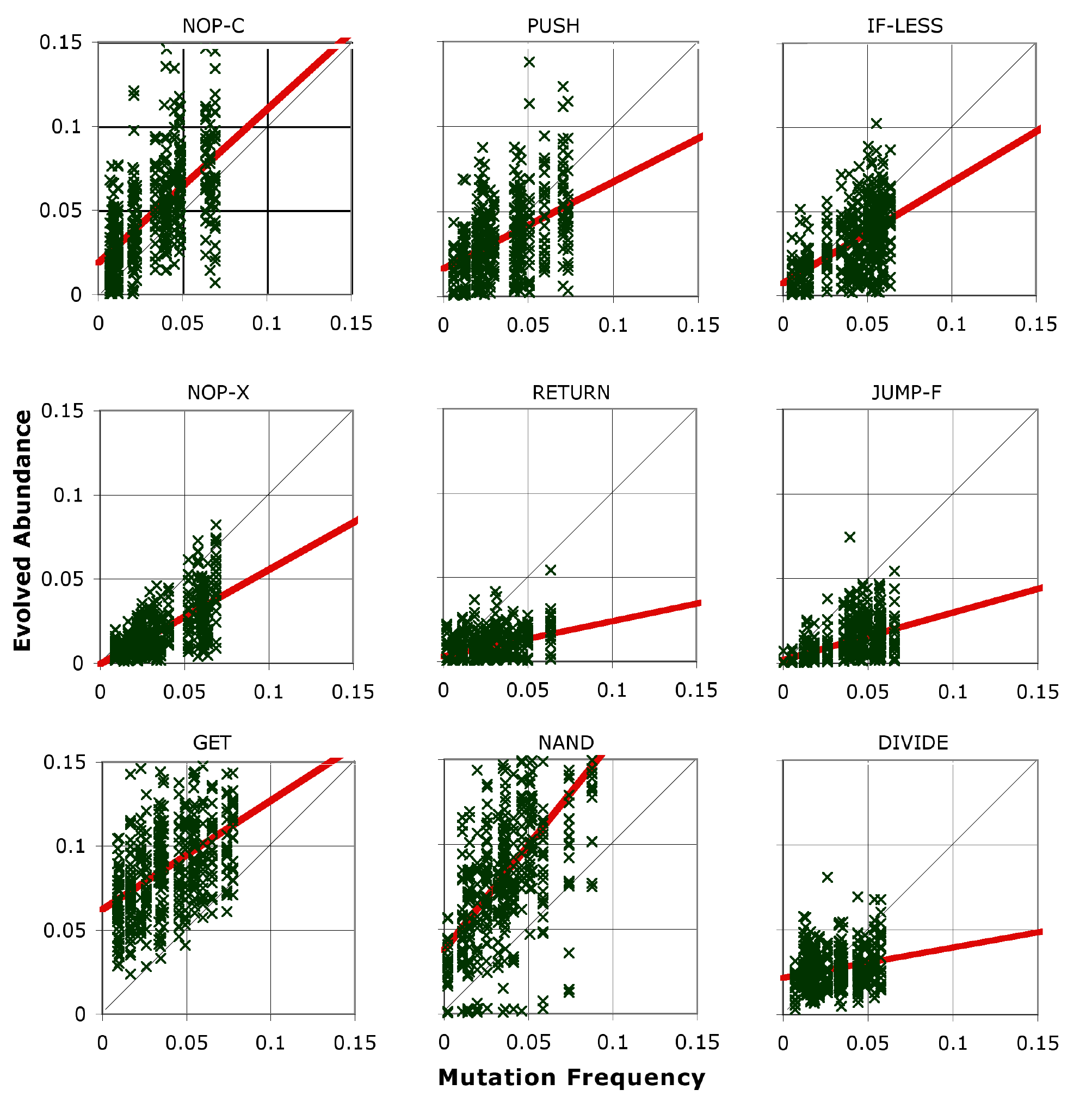} 
   \caption{The relationship between frequency of appearance in mutation and
abundance in the terminal population for nine instructions.   Trends above or below the
unity line (gray) indicate a tendency to be selected for or against, respectively.  GET typifies an 
instruction which frequently conveys a fitness advantage; it tends to be
incorporated into genomes at a high level regardless of how often it appears as
a mutation.  Conversely, RETURN is not necessary for basic functions and is often
lethal as a mutation, so it remains at a low level in nearly all runs, even when it
appears frequently as a mutation.   N=450, 25 runs each using 18 different 
mutation frequency profiles. Trend lines (red) are a least-squares linear fit. The grey line is the unbiased assumption where the mutation frequency equals the evolved abundance. Axes are chosen such that this unbiased trend line is always a diagonal.}
   \label{fig:fig4}
\end{figure}

Figure~\ref{fig:fig4} shows the relationships between mutation frequency and evolved
population frequency for several instructions (the average relationship for all instructions is shown in Table 1).  Each plot shows 450 data points, obtained from the 25 replicates of each of the 18 experiments where the mutational frequency
spectrum was determined as described in the Methods.  In each plot,
if selection did not constrain the organisms's use of each instruction,
we would expect the data to fall on or near the unity line, reflecting
that the organisms incorporated the instructions at the same rate at which they
appear in mutation (that is, the rate at which they are ``formed").  Data points above the unity line represent
populations that used an instruction at higher than the expected rate,
meaning that those instructions were preferentially incorporated into
the evolving genomes.  Points below the unity line represent
instructions that were selected against.  The slope of the distribution
represents the extent to which an instruction's appearance in the final
population depends on its availability in mutation.

While the data are widely distributed, some significant trends are
evident.  NOP-C, PUSH, and IF-LESS are close to neutral in average
adaptive utility: they show broad distributions near the unity line. 
All three of these instructions exist in parallel with other
instructions that can, to some degree, replace their function.  NOP-C is
part of a complementary set of address labels including NOP-A and NOP-B,
and an organism can construct labels using only a pair of NOP
instructions if one is not available.  PUSH is a stack operation, and
organisms can function largely using register (as opposed to stack)
storage if necessary.  The flow-control instruction IF-LESS can be
entirely replaced by IF-N-EQU and to some degree with IF-BIT-1, if it is
not available.

NOP-X, RETURN and JUMP-F are examples of maladaptive instructions that
are selected against, with most of their instances appearing below the
unity line.  NOP-X is a neutral but non-functional operation that merely
consumes a single CPU cycle.  While it does not harm an organism, if
inserted into a loop it can cause a significant delay in the time
required for a genome to complete tasks and reproduce.  We see it
somewhat selected against.  RETURN and JUMP-F, however, are flow control
instructions that are both unnecessary (their functions can be
completely replaced by other flow instructions like JUMP-B) and
generally maladaptive.  When a JUMP-F or RETURN is inserted or
substituted into a genome, the resulting genome will often skip large
blocks of instructions or enter infinite loops: such mutations are
usually fatal.  As a result, these are the two most strongly
selected-against instructions.

Obviously beneficial instructions include GET and NAND, which are
essential for the completion of mathematical tasks.  GET and PUT (not
shown in Fig.~\ref{fig:fig4}) are responsible for input and output within the computational
metabolism; genomes cannot gain any advantage over the ancestor without
using them.  As a consequence they are strongly selected for.  NAND is also present in
high abundance in nearly all populations, as it is used to perform
computations on the input stream accessed via GET (NAND is the only instruction available to compute logic tasks, and therefore must be used increasingly in order to achieve higher fitness).  However, it shows a
strong--in fact greater than unity--dependence on the input frequency of
mutation.  When NAND is produced more often, tasks evolve more quickly and as a consequence NAND instructions accumulate in the sequence, leading to an evolved abundance exceeding the production frequency.

DIVIDE is an interesting case: it has the least dependence on mutation
rate of any instruction, but even at near-zero mutation rate it appears
at a significant fraction of the population, shown by the high intercept
of the trend line relative to maladaptive instructions like NOP-X and
JUMP-F.  DIVIDE splits a genome in half, and is essential for the
reproduction of organisms after they have copied all of their
instructions.  However, if it appears in an inappropriate location the
organism will divide prematurely; this is nearly always fatal.  As such,
DIVIDE almost always appears exactly once per genome regardless of
mutation effects.  The vertical scatter seen in the DIVIDE frequency is largely
due to variation in the length of the evolved genomes.

Table 1 gives full results for the selection bias of the 28 instructions
we analyze.  Selection bias for each instruction is an approximation 
of the tendency of selection pressures to elevate (or suppress) the population 
concentration of a single instruction.   The selection bias $SB_{i}$ for instruction $i$
 is computed via:

\begin{equation}
SB_{i} = \sum_{n} C_{i,n} - \mu_{i ,n} \label{bias}\;,
\end{equation}
where $C_{i,n}$ is the fractional population concentration of instruction $i$
in trial $n$ and $\mu_{i,n}$ is the probability of instruction $i$ appearing 
via mutation in trial $n$.  This bias can be positive or negative.
The instructions in Table 1 are listed sorted by this 
measure (instructions with the strongest selection bias  are at the top). If an instruction is incorporated into the genome 
at same frequency at which it appears in mutation, we expect $SB$ to be zero. 
We also list the slope
of the linear fit for each instruction, which indicates the degree to which
an instruction's population abundance depends on its mutation frequency.
In order to test whether the linear fit can significantly distinguish the abundance distribution of the instruction from the unbiased assumption (the line with unit slope and vanishing intercept), we conducted a test of variances (F-test) and listed the value in Table 1, along with the associated P-value. According to this test, all instructions deviate significantly from the neutral evolution assumption.

\begin{table}
\caption{Selection biases [Eq.~(\ref{bias})], slope of evolved instruction frequency vs. mutation frequency (see Fig.~\ref{fig:fig4}), value of the F-test variable, and probability that this F-value could have been the result of chance (P-value), for the 
28 instructions we analyzed.}
\begin{tabular}{|c|c|c|c||c|c|}\hline
\multicolumn{2}{|c|}{Selection Bias  }&
{Slope $\pm$ std err.}&F & P-value\\
\hline
PUT  & 31.1 & 0.83 $\pm$ 0.05& 1,801 & 0\\ 
GET &  21.6 &0.65 $\pm$ 0.06& 847&  0\\
NAND & 20.7 & 1.24 $\pm$ 0.07&416&  0\\
NOP-B & 12.6 & 0.94 $\pm$ 0.07 &229& 0\\
NOP-C & 7.2 & 0.91 $\pm$ 0.06&93 &0\\
INC & 1.9 & 0.47 $\pm$ 0.04&80&0\\
SWAP & 1.4 & 0.89 $\pm$ 0.05&10.2&4.56$\times 10^{-5}$\\
ALLOCATE & 1.2 & 0.55 $\pm$ 0.03&159& 0\\
ADD & 0.2 & 0.75 $\pm$ 0.05& 12.4& 5.74$\times 10^{-6}$ \\
PUSH & -0.4 & 0.51 $\pm$ 0.05& 50& 0 \\
POP & -1.1 & 0.65 $\pm$ 0.05 & 33& 3.8$\times 10^{-14}$\\
DIVIDE & -1.2 &  0.18 $\pm$ 0.04& 285& 0\\
SHIFT-L & -1.6& 0.55 $\pm$ 0.03 &139&  0\\
SEARCH-F & -2.6  & 0.36 $\pm$ 0.04& 173 & 0\\
IF-N-EQU & -3.0  & 0.57 $\pm$ 0.03&144& 0\\
JUMP-B & -3.4 & 0.25 $\pm$ 0.02&731& 0\\
COPY & -3.6  & 0.30 $\pm$ 0.04 &231&  0\\
IF-LESS & -3.9 & 0.60 $\pm$ 0.05&95& 0 \\
DEC & -4.3 &0.62 $\pm$ 0.04&128& 0 \\
SUB & -4.8 & 0.71 $\pm$ 0.04&106&  0\\
CALL & -5.2 & 0.39 $\pm$ 0.06&132&0 \\
SWAP-STK & -7.0 & 0.45 $\pm$ 0.03 &450& 0 \\
RETURN & -7.4& 0.21 $\pm$ 0.02 &1562& 0\\
NOP-X & -7.6 & 0.56 $\pm$ 0.03& 539&0\\
IF-BIT-1 & -9.1  & 0.36 $\pm$ 0.03&940&0 \\
SEARCH-B & -10.2 & 0.33 $\pm$ 0.02&1,477& 0 \\
SHIFT-R & -10.4 & 0.48 $\pm$ 0.02& 1,137& 0\\
JUMP-F & -11.1 & 0.28 $\pm$ 0.02& 1,751& 0 \\
\hline
\end{tabular}
\end{table}

\section*{Discussion}

Artificial life is a useful tool for astrobiology, in that it can
examine the fundamental processes of life with an eye towards
identifying universal phenomena: features of life that may be
detectable regardless of a lifeform's substrate or particular form. 
 It may be seen, therefore, as an approach toward solving the
``single data point" problem, that is, that we know only one example
of evolved life (the terrestrial biosphere), and therefore cannot
draw scientific conclusions about the universality of features we
observe.  Using Artificial Life techniques, we can test conjectures about observable
invariants of life; other examples include measuring the reduction
of local entropy induced by cellular automata in an artificial
chemistry~\citep{CentlerF_etal:2003}.

We have demonstrated the repeatability of the MADB 
in populations of avidians, and characterized the signature's robustness with
respect to alterations in the underlying physics.  We find that although
significant variations in monomer abundance patterns do appear as
evolutionary experiments are repeated, general features (such as the
selection for mathematics instructions, and the suppression of
frequently-lethal flow-control instructions) are conserved.  This feature mirrors observations of functional and structural constraints 
on the composition of proteins in the terrestrian biosphere, where for example hydrophobicity or stability requirements constrain the type, but not the identity, of amino acids incorporated into proteins~\citep{Forsdyke2005}. 

More importantly, in no case does the evolved abundance pattern ever resemble
the pattern predicted by the system's physics, which is the most
important characteristic of a life-diagnostic or biosignature.  This
robustness clearly derives from evolutionary necessity. The organisms'
metabolism and composition are subject to selection for fitness, and the
features of that composition will be impressed upon their environment. This observation also has a parallel in terrestrian biochemistry, where  mutational constraints (giving rise to GC bias) can influence the amino acid composition of a protein~\citep{Guetal1998,SingerHickey2000}
but cannot change the basic pattern of relative residue abundances into those observed in abiotic samples. 

This conclusion indicates one possible direction for the search for
non-terrestrial life via a method that is agnostic of terrestrial biochemistry,
i.e., ``non-Earth-centric" life detection.  By modeling or recording the
range of plausible abiotic formation ratios of various chemical
compounds, we may examine samples for compounds appearing outside of
those ranges.  Measurements of chemical concentrations that deviate from
those ranges may indicate that an evolved metabolism is selectively
synthesizing useful compounds. On the contrary, because diagenesis can obscure the biosignature if enough time has passed since its deposition, the absence of a detectable signature is not necessarily conclusive for the absence of historical life.

It is important to recognize the distinction between the formation of the
MADB via biosynthesis and its detectability in an environmental sample.   
If the biotic signature was laid down long  before observation, 
diagenesis could obscure the MADB because it is possible that different
monomers degrade at different rates. For example, when natural 
sediments degrade over time, amino acids of low molecular weight 
can become predominant simply because they are more 
stable~\citep{AbelsonPH_HarePE:1969,ElsterH_etal:1991}. 
In addition, the signature will only be measurable
if the population of organisms is sufficient to generate measurable
biomass with respect to the background chemistry. Since we examine 
only the composition of the avidians themselves, this study is modeling
only the formation and evolution of the signature, not whether it would
be measurable in the environment.   We refer the reader to our prior 
work~\citep{DornED_etal:2011} for further discussion of this issue.

Understandably, there are considerable practical limitations and obstacles to determining the baseline distribution to which we would compare an observed pattern of molecules. For example, the ability of this strategy to reject false positives
depends critically on our ability to thoroughly characterize the range
of possible abiotic distributions in advance. Such an endeavour
should begin with the most exhaustive experimental analysis possible of
monomer formation in all conceivable abiotic conditions, and we should
remain vigilant to the possibility that unconsidered special cases (i.e.,
unusual combinations of local environmental chemistry, temperature,
radiation, or other factors) might include dynamics that produce an
unexpected abundance pattern. While theoretical modeling and numerical simulation of planetary atmospheres and surface chemistry are important components of establishing the baseline, if possible they should be supplemented by simulations in a laboratory. For example, the geochemistry of exoplanets may be modeled using specific geochemical cycles to constrain spectral signatures~\citep{KalteneggerSasselov2010}, while Mars's evaporite geochemistry is readily simulated under Martian environmental conditions in the laboratory~\citep{Mooreetal2010}. In some instances the knowledge of the background distribution does not have to be precise, such as when abiotic chemistry predicts a smooth distribution of polymer abundances while the biotic distribution is discrete (cf. the Lego principle of \cite{McKay:2004,DaviesPCW_etal:2009}). Once a candidate set of polymers has been identified and a baseline distribution suggested, planetary missions could be designed that measure thousands of chemical relative abundances using for example on-chip liquid chromatography, or other methods such as Raman spectroscopy or UV fluoresecence~\citep{McKay2010}. However, the technology to perform high-throughput targeted relative abundance measurements is probably still years away. Given such measurements, however, standard machine-learning techniques can be applied to distinguish biotic from abiotic patterns~\citep{DornED_etal:2003,Dorn2005}.

Yet, even in the light of such practical difficulties,  we see the MADB as an important biosignature for life
detection, worthy of investigation because it is the inevitable result
of a fundamental life process (evolutionary selection) and is completely
independent of information about any specific biochemistry.  If the
abiotic distribution is characterizable, the MADB should be detectable as
long as the lifeform under study employs in its metabolism any members
of the chemical family under study.

\section*{Acknowledgments}
We would like to thank Ken Nealson for discussions and collaboration in an earlier phase of this research. 
This material is based in part upon work supported by the National Science Foundation under Cooperative Agreements No. DEB-9981397 and No. DBI-0939454. Any opinions, findings, and conclusions or recommendations expressed in this material are those of the authors and do not necessarily reflect the views of the National Science Foundation.

\section*{Author Disclosure Statement}
The authors state that no competing financial interests exist.
\bibliography{biomarkers,alife}
\bibliographystyle{apalike}
 \end{document}